\newcommand{\grouphead}[2]{%
  \addlinespace[2pt]
  \multicolumn{2}{@{}l@{}}{\cellcolor[gray]{0.92}\parbox[c][14pt][c]{\linewidth}{\ \textbf{#1}}}\\
  \multicolumn{2}{@{}p{\linewidth}@{}}{\itshape\small #2}\\
  \addlinespace[3pt]

}
\documentclass[manuscript]{acmart}
\AtBeginDocument{%
  }

\setcopyright{acmlicensed}
\copyrightyear{2018}
\acmYear{2018}
\acmDOI{XXXXXXX.XXXXXXX}
\acmConference[Conference acronym 'XX]{Make sure to enter the correct
  conference title from your rights confirmation email}{June 03--05,
  2018}{Woodstock, NY}
\acmISBN{978-1-4503-XXXX-X/2018/06}

\usepackage[table]{xcolor}
\usepackage{longtable}
\usepackage{pdflscape}   
\usepackage{multirow}
\usepackage{booktabs}
\usepackage{booktabs}
\usepackage{colortbl}
\begin{document}

\title{``A Necessary Evil'': Teenagers' Sensemaking of Privacy and Safety Settings on Social Media}
\author{Jingxin Dong}
\email{dong11@iu.edu}
\orcid{0009-0003-8699-6776}
\affiliation{
  \institution{Indiana University}
  \city{Bloomington}
  \state{Indiana}
  \country{USA}
}

\author{Lingyun Chen}
\email{lch2@iu.edu}
\orcid{0009-0007-9191-0964}
\affiliation{
  \institution{Indiana University}
  \city{Bloomington}
  \state{IN}
  \country{USA}
}

\author{Chen Ling}
\email{ccling@iu.edu}
\orcid{0000-0001-5082-7557}
\affiliation{
  \institution{Indiana University}
  \city{Bloomington}
  \state{Indiana}
  \country{USA}
}

\author{Colin M. Gray}
\email{comgray@iu.edu}
\orcid{0000-0002-7307-1550}
\authornote{Colin has engaged in paid expert witness and consulting work on cases relating to social media. This expert witness work is independent from the data and methods presented in this paper.}
\affiliation{
  \institution{Indiana University}
  \city{Bloomington}
  \state{Indiana}
  \country{USA}
}
\begin{abstract} 
Social media platforms are embedded in teenagers' daily lives, supporting friendship and identity while exposing teenagers to unwanted contact and privacy harms. Previous scholarship has documented how attention capture strategies and dark patterns shape social media use, and we extend this work to better understand platform settings that ostensibly provide privacy and safety protection. We report on think-aloud sessions with 11 teenagers aged 14 to 17 who completed six privacy and safety tasks on Instagram, TikTok, Snapchat, and YouTube. We show how participants worked out what a setting meant through their routines, boundaries, and prior experiences, how they accommodated protections softer and less predictable than expected, and how they treated the platform as the authority on what protection should look like. We argue that feature-by-feature evaluation cannot establish whether teenagers are protected, and that platforms should carry the obligation to show that a protective action took effect and is durable.
\end{abstract}



\received{20 February 2007}
\received[revised]{12 March 2009}
\received[accepted]{5 June 2009}

\maketitle

\section{Introduction}
In March 2026, a Los Angeles County Superior Court jury found Meta and Google negligent for the design of Instagram and YouTube and for their failure to warn young users, deciding the first bellwether trial in California's coordinated Social Media Cases (JCCP 5255) \citep{allyn2026meta}. Jurors connected the design of the two products to mental health harms that the plaintiff carried from childhood through adolescence, and the decision pulled public argument toward how platform design shapes adolescent wellbeing \citep{allyn2026meta}. Verdicts of this kind sharpen the concern that drives this paper, which is whether social media privacy and safety controls give teenagers meaningful protection and agency. Social media occupies a complex place in teenagers' lives, supporting friendship, communication, and identity formation \citep{Christoferson2016-rx, George2015-ep} while also exposing them to unwanted contact, privacy violations, manipulative design, and patterns of use that can be difficult to regulate \citep{Ali2024-xn, Behfar2024-mw, Chen2024-hm, Gairola2026-do}. The CDC's nationally representative 2023 Youth Risk Behavior Survey found that 77\% of U.S. high school students used social media at least several times a day, and frequent users reported higher prevalence of electronic bullying and persistent feelings of sadness or hopelessness than less frequent users \citep{Young2024SocialMediaBullying}. Social media is woven into adolescent daily life at a scale that survey work documents clearly, even where causal questions remain open.

Adolescence is a formative period for developing social boundaries, personal autonomy, and self-regulation, so protections built into social media interfaces may shape how teenagers recognize and respond to risk \citep{Wisniewski2018-yp, Mondal2019-jw}. Platforms present privacy and safety settings as tools for making participation safer and more manageable. Nevertheless, making controls available does not ensure that teenagers understand their purposes, trust how they operate, or experience them as meaningful protection \citep{Dong2026-ip, Schafer2024-gu}. Examining teenagers' own interpretations is therefore essential to understanding whether platform controls support agency in everyday social media use \citep{Wan2025-yh, Park2025-jg, Zhao2022-uu}.

Prior research shows that privacy and safety controls can themselves introduce barriers to meaningful user control \citep{Schaffner2022-af, Dong2026-ip}. Within human-computer interaction (HCI), Gray and colleagues extended Brignull's account of dark patterns by showing how recurring interface strategies can privilege organizational goals over user agency \citep{BrignullDarkPatterns, Gray2018-it}. Protective controls may obstruct action, use ambiguous language, or make privacy choices easy to reverse \citep{Gray2024-rf, Schaffner2022-af}, while subtle manipulation can remain difficult for users to recognize \citep{Mildner2023-px, Schafer2024-gu}. Federal oversight remains fragmented across agencies that hold different mandates \citep{FTC2022DarkPatterns}, as the Federal Trade Commission (FTC) addresses deceptive design and consumer privacy, the Department of Justice (DOJ) handles antitrust and criminal enforcement, and the Federal Communications Commission (FCC) exercises narrower authority over communications services \citep{FCCOutsideJurisdiction}. Beyond design and regulation, Palen and Dourish describe privacy as an ongoing negotiation of disclosure and interpersonal boundaries \citep{Palen2003-hu}, while research with teenagers shows that settings are configured around audiences, social needs, and unwanted contact \citep{Wan2025-yh, Park2025-jg, Zhao2022-uu}. Research on adolescent online safety further emphasizes autonomy and resilience \citep{Wisniewski2018-yp}, while studies of privacy preferences show that teenagers' decisions change with experience and circumstance \citep{Mondal2019-jw}. Ackerman's sociotechnical gap connects the literature by distinguishing what platforms can technically provide from what teenagers require socially \citep{Ackerman01092000}. Our study examines how teenagers interpret a broader range of privacy and safety controls during use and relate expected protections to their routines, boundaries, and prior experiences.

We examine what participants noticed, how they understood each control, what outcomes they expected, and whether the control could support their routines, boundaries, and responsibilities. We conducted moderated sessions using a think-aloud protocol with 11 teenagers aged 14 to 17 as they completed six tasks involving notifications, screen time, account privacy, reporting, data access, and account deletion on social media platforms they already used. We analyzed the sessions through reflexive thematic analysis and developed two themes. \textit{Living With and Around Platform Design} describes how participants worked out what a control meant and whether it was useful through their own routines, boundaries, and prior experiences, and they often accommodated settings that were softer or less predictable than they expected. \textit{Teens Defer to the Platform as Authority} captures how dense or unclear settings were often left unchanged, how extra steps could be interpreted as evidence of safety, responsibility, or legitimate design, and how repeated use and limited alternatives could strengthen reliance on the platform. Both themes carry the sense of the platform as a \textit{necessary evil}, where participants recognized individual controls as weak, partial, or confusing and still relied on platforms they perceived as familiar, authoritative, and difficult to avoid. Across both themes, platform authority develops through repeated encounters between interface cues and participants' routines, relationships, and prior experiences, and it does not operate as a separate starting assumption. Our study is guided by the following research question: \textit{\textbf{How do teenagers interpret the meaning and effectiveness of social media privacy and safety controls, and how do they relate these interpretations to their everyday practices, boundaries, and responsibilities?}}

Our contributions are threefold. First, we document what participants noticed in privacy and safety controls during situated use, how they judged the strength and purpose of those controls, and how interface cues met their routines, boundaries, and prior experiences across six tasks. Second, we show that teenagers can recognize a control as weak and still treat the platform as the authority on what protection should look like, and that the friction surrounding a protective flow can itself become the reason they trust it, which complicates design responses aimed at making manipulative patterns more visible. Third, we argue that evaluating controls one feature at a time cannot establish whether teenagers are protected, and we propose an asymmetric allocation of responsibility in which platforms carry the obligation to show that a protective action took effect and holds without continued management.

\section{Related Work}

\subsection{Platforms, Attention Capture Strategies, and Dark Patterns}
Many social media platforms rely heavily on advertising-based business models that depend on sustained user attention and engagement \cite{parisi2022will, goldhaber1997attention}. The design consequences of that dependence have proven easier to demonstrate than the popular language of addiction used to describe them. Advertising-supported platforms benefit from repeated and prolonged engagement, which they sustain through personalized feeds, recommendations, autoplay, notifications, and continuously available content that make another item or social interaction immediately available \citep{Gray2026-ks}. Researchers have consequently described difficult to control engagement through terms such as problematic, compulsive, and ``addictive'' use, though the components of addiction can be difficult to cleanly separate in some cases 
\citep{Fournier2023-ad}. HCI scholarship addresses that ambiguity by examining how interface design supports or disrupts users' intentions, and studies that trace agency to specific features show that a single platform can both undermine and support deliberate use. Lukoff and colleagues found that autoplay and recommendations on YouTube weakened participants' sense of agency, while search and playlists supported more deliberate viewing \citep{Lukoff2021-yi}. Work on intentional Twitter use further confirms the asymmetry from the opposite direction, where feed filters and indicators showing that users had exhausted new content improved agency, while usage dashboards and prompts to close the application did not \citep{Zhang2022-bg}. Both studies locate the difference in particular interface features and not in the user, which turns the question from why people cannot stop into what the interface makes easy.

Attention capture research answers that question by specifying the mechanisms, moving from the general observation that platforms hold attention to an inventory of the designs that support attention capture. In a review of 43 papers, Monge Roffarello and colleagues identify eleven attention capture damaging patterns that exploit psychological vulnerabilities through deception or seduction \citep{Monge-Roffarello2023-qc}. Infinite Scroll continuously loads content and Neverending Autoplay begins another video without requiring a new choice. The pull-to-refresh gesture pairs a familiar action with an uncertain prospect of appealing content, drawing on the variable reward logic associated with repeated checking, while Guilty Pleasure Recommendations use individualized temptations and Recapture Notifications prompt another session after users have left. Mildner and colleagues sort comparable features into engaging strategies that keep users occupied, such as interactive hooks and social brokering, and governing strategies that shape decision-making within a service, such as decision uncertainty, labyrinthine navigation, and redirective conditions \citep{Mildner2023-yw}. A structured content analysis of seventeen Very Large Online Platforms, conducted from a teen account perspective, extends the inventory to the platform level and identifies 63 designs and 583 instances organized around pressuring, enticing, trapping, and lulling users into continued engagement \citep{Chen2024-hm}.

Attention capture research establishes what engagement designs do, and dark pattern research asks when doing it counts as a harm to users, which shifts the question from commercial effectiveness to the conditions under which an interface overrides a person's own goals. Work in this area has moved from catalogs of individual ``tricks'' to accounts of the business models that sustain them \citep{Gray2023-ay}. Brignull, along with Conti and Sobiesk, described interfaces that deliberately privilege designer goals over user experience \citep{Brignull2023-rs,Conti2010-zf}. Gray and colleagues organized practitioner examples into Nagging, Obstruction, Sneaking, Interface Interference, and Forced Action, which gave HCI a shared set of terms for recurring manipulative strategies \citep{Gray2018-it}. Privacy research extended that account by showing how defaults, complexity, and information practices can undermine disclosure choices and consent \citep{Bosch2016-ei,gray2021dark,kelly2023documenting}, and a large audit documenting 1,818 dark pattern instances established that manipulative designs operate as repeated commercial practices \citep{mathur2019dark}. Within social media, account deletion research reveals obstruction, confusing terminology, and inconsistent options that can delay or discourage exit \citep{Schaffner2022-af}, and studies of feed settings show that perceived control can diverge from actual system behavior, although the strength of placebo effects varies across designs and users \citep{Vaccaro2018-yx,Hsu2025-ag}. Multilevel frameworks then connect these patterns to the systems in which they operate. Gray and colleagues consolidate ten academic and regulatory taxonomies into a three level ontology that relates observable interface features to broader strategies, where Attention Capture appears at the meso-level under Forced Action and engaging strategies map onto Attention Capture and Social Engineering while governing strategies map onto Interface Interference and Obstruction \citep{Gray2024-rf}. The Dark Patterns Knowledge Stack builds on that ontology and aligns the sociotechnical landscape, goals and intentions, the interface, and users' limitations and knowledge, which connects advertising incentives and engagement metrics with the interface mechanisms described above and with the vulnerabilities that shape users' encounters \citep{Gray2026-ks}. Empirical work on recognition reports uneven results, since regular social media users can distinguish manipulated interfaces under structured evaluation \citep{Mildner2023-px}, while youth studies show partial awareness in which defaults and visually subtle patterns often remain harder to identify \citep{Schafer2024-gu,Schafer2025-xw}. Recognition alone does not determine response, since family experience, peer practices, and prior knowledge shape how teenagers interpret risks and develop resistance \citep{Sanchez-Chamorro2024-hv,Kelly2025-re}, and social and technical circumstances mediate whether resistance is available at all \citep{Sanchez-Chamorro2025-sk}. Closest to the present work, an expert evaluation of teen accounts on four social media platforms found dark patterns in every privacy and safety task on at least one service, with account deletion and notification management creating especially substantial friction \citep{Dong2026-ip}. In the present paper, dark pattern research describes the platform environments in which privacy and safety controls appear, and it does not supply the analytic frame for our findings.

\subsection{Usable Privacy, Sensemaking, and Self-Limiting Tools}
Usable privacy research has established that finding and operating a control settles only part of the question, because its criteria stop at the point of action and says little about what users believe the action accomplished. Habib and Cranor define privacy choice usability through criteria spanning user needs, awareness, comprehension, ability and effort, sentiment, decision reversal, and nudging patterns, extending the standard beyond task completion to whether people can locate a choice and understand its implications \citep{habib2022evaluating}. An audit of 150 websites shows how often that standard fails in practice, with privacy choices appearing in inconsistent locations, frequently omitting information needed to understand an action, and sometimes linking to pages lacking the promised option \citep{Habib2019-dl}. Presentation carries its own influence, and research on privacy and security nudges raises the question of whose goals a design should serve \citep{Acquisti2017-nu}. However, the usability frame reaches its limit where operation is easy and consequence is opaque. Vaccaro and colleagues found that functional and randomly implemented feed controls both increased perceived control, though a later replication found a smaller satisfaction effect \citep{Vaccaro2018-yx,Hsu2025-ag}. A control can therefore satisfy every usability criterion available and still leave users unable to say what it did.

Sensemaking research explains what users do with that residue, showing interpretation to be ongoing, revisable, and frequently mistaken. Eslami and colleagues identified ten theories of Facebook feed curation and found that a probe exposing alternative feed outputs helped participants develop explanations informing subsequent plans \citep{Eslami2016-hn}. DeVito extends the account by showing how users revise theories as platforms change and how perceived platform spirit conditions the process \citep{DeVito2021-pv}. Controls supply another cue within such reasoning, and conflicting understandings can persist around a single visible and operable control, as when Instagram users approached the Not Interested feature with varied expectations about the people and content their feedback would affect, then struggled to attribute later feed changes to a particular action \citep{Hong2025-pc}. Moreover, privacy compounds the interpretive problem because its object is social and not computational. Palen and Dourish conceptualize privacy as a dynamic boundary regulation process organized through ongoing negotiations over disclosure and access \citep{Palen2003-hu}, and on social media those boundaries extend through other people, who can redistribute information and alter the contexts in which it becomes visible \citep{Marwick2014-uj}. Teenagers manage this across personal and interpersonal levels. Specifically, De Wolf distinguishes management of personal disclosures from coordination around shared information, while Chou and Chou show that proactive prevention and reactive recovery rely on different threat appraisals and forms of efficacy \citep{De-Wolf2020-me,Chou2023-ru}. In practice, teenagers on Instagram combine built in controls such as Close Friends with secondary accounts and offline coordination, and learn about available features from peers \citep{Zhao2022-uu}. Research on feeds and on disclosure is well developed, and it remains thin on how users read the controls themselves.

Research on self-limiting tools carries the same interpretive question into the domain where users act on their own behalf, and it consistently finds that a tool's value depends on whether what it enforces matches the user's own definition of unwanted use. A review of digital self control tools found blocking and interface removal to be the dominant interventions \citep{Lyngs2019-sc}, while a later analysis of user reviews showed people seeking controls strong enough to interrupt unwanted use and flexible enough to match personal definitions of distraction \citep{Lyngs2022-gl}. Fit with a user's own goals depends on the situation, since desired control on YouTube varied with whether users arrived with a specific intention \citep{Lukoff2021-yi}. Chirp bears this out, comparing external supports such as a usage dashboard and prompts to close the application against internal supports built into the feed itself, and only the internal supports increased agency, while usage time held steady across conditions \citep{Zhang2022-bg}. \textit{SwitchTube} operationalized changing intentions through Focus and Explore modes, and access to selectable modes improved agency, satisfaction, and alignment with viewing goals compared with an interface centered on recommendations \citep{Lukoff2023-xz}. Platform-provided controls behave the same way, since Netflix viewers used planning, episode limits, and alarms, yet reported that autoplay and weak stopping cues extended viewing beyond their intentions \citep{Schaffner2023-nf}, and a randomized experiment found that disabling autoplay reduced daily playback and session duration while leaving participants divided over restoring it, some valuing convenience and others the added moment for reflection \citep{Schaffner2025-ap}. Each of these bodies of work studies the controls with adults, and as objects whose usability or efficacy can be measured from outside. Usable privacy research asks whether a person can complete a privacy action, and research on self-limiting tools asks whether a tool changes behavior. We ask what teenagers believe these controls could be doing for them.

\subsection{Youth Expectations, Belonging, Identity, and Trust}
Teenagers form expectations of social media from inside systems they have always known, which gives platform conventions a practical authority that does not depend on approval of the companies behind them. Imagined affordances describe how those expectations emerge among users' perceptions and attitudes, technologies' materiality and functionality, and designers' intentions \citep{Nagy2015-fl}. Foundational research on networked publics locates the same process in adolescent social life, where young people gather, socialize, and contribute to peer culture while negotiating persistent content and partially invisible audiences \citep{boyd2008-np}. Patterns of social media use have also changed over time, including major disruptors such as the COVID-19 pandemic, which may have increased reliance on these platforms \citep{Liang2023-pd}.

Belonging and identity give platform participation its value for teenagers. Livingstone connects teenagers' profile creation and content sharing to intimacy, privacy, and self-expression, treating risk and opportunity as two sides of the same practice \citep{Livingstone2008-ro}. Davis locates the mechanism in friendship itself, finding that digital communication supported belonging and self disclosure while offline differences in age and gender continued to shape online exchanges \citep{Davis2012-fr}. Adolescents' own accounts show the same priority at scale, with attempts to connect forming the most common purpose across posts and varying in frequency by platform \citep{Alluhidan2024-yz}. Further, anticipated connection can outweigh the content itself, since teenagers returned to Instagram expecting social contact even when much of the feed felt uninteresting \citep{Landesman2024-vs}. Limits on visibility, sharing, or use therefore alter opportunities for inclusion, and identity work raises the stakes further as young people present themselves to overlapping audiences and read peer responses as information about recognition. Privacy and self-limiting controls carry social consequences when their use changes how a teenager appears to friends or participates in group routines.

Trust runs along two tracks that youth research keeps separate, one toward the people a teenager is connected to and one toward the company hosting the connection. A systematic review distinguishes trust in information, other users, and platforms, and identifies different dimensions used to characterize each target \citep{Zhang2023-ni}. Kim and colleagues locate the first track in relationships, finding that teenagers sought casual sharing to strengthen close ties while ambiguous audiences and unforgiving norms discouraged disclosure \citep{Kim2025-xm}. On the other hand, youth are twice as likely to distrust as to trust social media companies' ability to produce a fair resolution after online harassment, with 41 percent distrusting against 20 percent trusting, and some respondents linked inadequate action to revenue motives \citep{Schoenebeck2021-yt}. Distrust of the company does not translate into withdrawal from the platform, and trust research finds the two coexisting, since users evaluate different aspects of the same system and can hold confidence in one while doubting another \citep{Zhang2022-ky,Zhang2024-ln}. Workshops with young people found suspicion of corporate tracking alongside a perceived obligation to accept platform terms and retain a presence for social relationships, and participants often adopted permissive privacy practices to avoid disrupting ordinary interaction even while recognizing the reuse of personal data \citep{Pangrazio2018-zx}. Reliance of that kind grants platforms practical authority by allowing their defaults, categories, and participation rules to structure access to peers. Draper and Turow name the resulting condition digital resignation, where concern persists alongside a felt inability to change institutional practice \citep{Draper2019-ip}, and Hargittai and Marwick describe a related apathy in which users doubt that any available action would protect them \citep{Hargittai2016-xn}. Resignation implies a lowered expectation, which leaves open how to read young people who keep expecting protection from a platform they already doubt. Familiarity leaves scrutiny limited, since in a recognition task with fifth graders, only three of 51 noticed a malicious privacy default \citep{Schafer2024-gu}, while family and peers shaped what teenagers noticed about manipulative design and which coping strategies they treated as available \citep{Sanchez-Chamorro2024-hv}. Youth trust in these accounts is measured as an attitude reported in surveys and workshops, at a distance from the moment a teenager is deciding whether a particular setting will hold.

\section{Method}

This paper examines how participants interpreted privacy and safety settings while completing common tasks on platforms they already used. We used one-on-one think-aloud interviews to capture what participants understood a feature to be for, how they expected it to work, and whether they would use it themselves.

\subsection{Participants and Recruitment}
To examine how teenagers made sense of privacy and safety settings in context, we recruited 11 teenagers (PS\_1–PS\_11), aged 14 to 17 (\(M = 15.6\), \(SD = 0.8\)), for one-on-one think-aloud interviews conducted over Zoom. Participant demographics and prior social media experience are summarized in Table~\ref{tab:participant_demographics}. We frame this as an interpretive, task-based interview study. Our goal was to understand how participants interpreted privacy and safety controls as they encountered them in situated interaction. We focused on tasks such as screen time, notifications, setting an account to private, reporting, downloading data, and account deletion because prior work has connected these settings to teen social media use, privacy management, platform friction, and dark pattern related experiences~\cite{Mildner2023-px, Gruzd2018-zu, Fahlman2018-kw, Schaffner2022-af, Zhang2022-bg, Pangrazio2018-zx, Hong2025-pc}. We used a think-aloud method to capture participants' in-the-moment interpretations, expectations, uncertainties, and justifications while they attempted to complete privacy and safety tasks~\cite{Ramey2006-ca,Nielsen2002-ai}. The approach allowed us to examine not only whether participants could find or complete a setting, but also what they believed the setting meant, how they expected it to work, whether they trusted it, and whether they imagined using it in their own everyday social media practices.

Before the full study, we conducted two pilot interviews to test the interview protocol and the think-aloud process on participants' personal phones. We recruited two young adults, a 19-year-old female and a 22-year-old female, close enough in age to our target range to approximate how teenagers would encounter the tasks. The pilots helped us refine the task wording, interview order, and follow up questions. We excluded both sessions from the analysis because they are outside our target user scope. We recruited participants for the full study through a standing participant registry maintained by one member of the research team, in which parents have consented to be contacted about research involving their children. We distributed a screening questionnaire to registry parents to establish eligibility and to collect background information about their teen's social media use, including which of TikTok, Instagram, Snapchat, and YouTube their teen used most often. Parents who were interested enrolled their teen in the study. We also used snowball sampling through participants' own networks to reach additional teenagers. We used the screening responses to assign each participant two platforms for the task session, selected from those the participant already used, so that the tasks were situated within existing platform familiarity.

\subsection{Research Ethics}
This study was approved by our Institutional Review Board (IRB). Before each interview, both the parent or guardian and the participant received digital consent and assent forms through Adobe Sign. These forms were signed before scheduling the session. At the beginning of each interview, we explained the study purpose, voluntary participation, confidentiality measures, and the right to skip any question or withdraw at any time. We also obtained permission to record the session. We also asked whether parents were present during the interview so participants could speak freely. In one case, a parent (PS\_07) chose to remain in the room during the session, which may have shaped the participant's responses.

\subsection{Teen Interview Study}
\begin{table}[t]
\label{tab:participant-demographics}
\centering
\footnotesize
\setlength{\tabcolsep}{3pt}
\renewcommand{\arraystretch}{1.1}
\rowcolors{2}{gray!12}{white}
\begin{tabular}{p{0.10\columnwidth} p{0.18\columnwidth} p{0.06\columnwidth} p{0.05\columnwidth} p{0.30\columnwidth} p{0.19\columnwidth}}
\toprule
\rowcolor{gray!25}
\textbf{Participant ID} & \textbf{Platform Performed during Interview} & \textbf{Gender} & \textbf{Age}  & \textbf{Platforms Used the Most} & \textbf{Time of First Use} \\
\midrule
PS\_01  & TikTok, Snapchat    & M  & 16 & TikTok, Instagram, Snapchat, YouTube & 2.5 years ago \\
PS\_02  & YouTube, Instagram  & F & 16 & TikTok, Instagram, Snapchat, YouTube & About 2--3 years ago \\
PS\_03  & Snapchat, Instagram & M & 15 & Snapchat & 2 years ago \\
PS\_04  & TikTok, YouTube     & M & 14 & TikTok, Instagram, Snapchat, YouTube & 1 year ago \\
PS\_05  & TikTok, Instagram   & F & 16 & TikTok, Instagram, Snapchat, YouTube & 2 years ago \\
PS\_06  & Snapchat, TikTok    & M & 17 & TikTok, Instagram, Snapchat, YouTube & A few years ago \\
PS\_07  & Instagram, TikTok   & M & 16 & TikTok, Instagram, Snapchat & About 1 year ago \\
PS\_08  & TikTok, Instagram   & F & 15 & TikTok, Instagram, Snapchat, YouTube & About 6 months ago \\
PS\_09  & Snapchat, TikTok    & M & 16 & TikTok, Instagram, Snapchat, YouTube & 5 years ago \\
PS\_10 & YouTube, Instagram  & F & 16 & TikTok, Instagram, Snapchat, YouTube & 4 years ago \\
PS\_11 & Instagram, YouTube  & M & 15 & TikTok, Instagram, Snapchat, YouTube & About 1--2 years ago \\
\bottomrule
\end{tabular}
\caption{Participant ID and study related background for the 11 participants in our sample, including the platforms used during the interview tasks, reported platform use, and time of first use.}
  \Description{Participant demographics and platform use for the 11 participants, one row each. Columns give participant identifier, the two platforms used during the interview tasks, gender, age, platforms used most, and time of first use. Ages range from 14 to 17, with six participants aged 16. Six participants are male and five female. Nine of the eleven report using all four of TikTok, Instagram, Snapchat, and YouTube; one reports Snapchat alone and one omits YouTube. Time of first use ranges from about six months to five years, with most reporting one to three years.}

\label{tab:participant_demographics}
\end{table}

\subsubsection{Interview Procedure}
The interview protocol had three main parts, covering platform familiarity, a think-aloud task session, and follow up questions. (1) We asked participants about their familiarity with each platform, including how long they had used it, how often they used it, and how well they understood privacy and safety settings. We also asked broader questions about privacy and safety online. The warm up questions established participants' existing routines, concerns, and prior experiences before the tasks, which we drew on when interpreting what they said during them. (2) We conducted a think-aloud task session in which participants completed privacy and safety tasks on two study platforms. The tasks included managing notifications, setting screen time limits, setting an account to private, reporting content, downloading data, and deleting accounts. During the tasks, participants were encouraged to narrate out loud what they were doing, what they noticed, what they thought a feature meant, and how they felt at each step. The interviews were conducted on Zoom, so participants shared their phone screens during the tasks. Screen sharing let us see how they navigated the settings in real time, including where they hesitated, where they got stuck, where they considered giving up, and where they were detoured to other parts of the interface. (3) We asked follow up questions about which tasks felt easy or difficult, what they expected from the platform, and what they thought about the overall experience. Some participants made their interpretation explicit during a task, while others did so only after attempting it, and the follow ups gave them room to say whether a setting felt useful, trustworthy, confusing, or worth using. One interview (PS\_04) was co-facilitated with a faculty advisor, Three interviews (PS\_01, PS\_02, PS\_03) included another researcher who helped take notes during the session, and the lead researcher wrote a short memo. When multiple researchers were present, the authors also discussed key observations together after the interview.

\subsubsection{Preparing the Corpus}
Our analysis focused on sensemaking, meaning how participants interpreted privacy and safety features while performing the tasks and reflecting on them during the interview. We operationalized sensemaking as any moment when participants interpreted what a feature was or what it was for, described how they expected it to work, evaluated whether they believed it would work, connected it to their own social media use, or inferred why the platform had designed or presented a control in a particular way. The purpose of this stage was to construct a corpus of interpretive material for thematic analysis. We screened all transcripts for segments containing interpretive content. We included excerpts in which participants explained the meaning or expected operation of a feature, drew on prior experience to understand it, compared it with another platform, connected it to their own routines or concerns, or reflected on what the platform appeared to be asking or encouraging them to do. We excluded segments that only narrated navigation, such as reading a button label or stating a menu location without interpretation or evaluation. Each included excerpt was then tagged with the platform and task in which it occurred, following Saldaña's structural coding \citep{Saldana2012-mo}. The structural tags did not represent themes and instead preserved the situated context of each interpretation so that we could examine how participants read the same category of control across TikTok, Instagram, Snapchat, and YouTube, which is the comparison reported in Section~4.1.1.

\subsubsection{Organizing Sensemaking Excerpts}
We next applied a second layer of nonexclusive analytic codes to organize recurring forms of sensemaking. The codes were Features, Lived Experience, Expectation, Data and Privacy, Usage, Comparison, and Reflection, and they functioned as retrieval and comparison tools instead of candidate themes. A single excerpt could receive several codes when participants moved among different forms of interpretation within the same moment. For example, a participant configuring screen time could simultaneously describe how the feature appeared to work, compare it with another platform, and connect the expected limit to their own daily routine. We also recorded whether an interpretation emerged without an interviewer probe directing the participant toward that interpretation (\textit{Organic}) or following a probe, clarification, assistance, or request for further explanation (\textit{Prompted}). The markers preserved information about how an interpretation entered the interview record and were considered when developing and wording thematic claims. We did not treat them as task outcomes or as measures of stronger or weaker evidence. The structural and analytic codes therefore served different purposes. Platform and task codes preserved where an interpretation occurred, while the analytic codes described what participants were making sense of within that encounter. Neither layer determined the final themes. Table~\ref{tab:codebook} in the Appendix gives the full definition of each structural and analytic code, a representative excerpt, and the number of applications.

\subsubsection{Developing Themes}
We then conducted an iterative reflexive thematic analysis informed by Braun and Clarke \citep{braun2022thematic}. We compared tagged excerpts within and across participants, tasks, platforms, and interpretation contexts, while repeatedly returning to the complete transcripts and interview memos when the surrounding interaction was necessary for interpreting an excerpt. We wrote analytic memos about recurring relationships, contradictions, and cases that complicated emerging interpretations, and we discussed candidate patterns among the research team and revised them through repeated comparison with the dataset. The themes reported below were constructed through interpretation of relationships among tagged excerpts, and they were not derived from tag frequencies or by combining individual codes into higher-level categories. We report the tag application counts in Table~\ref{tab:codebook} to describe the composition of the corpus and not to evidence the strength of any theme, and we did not assess intercoder reliability, since reflexive thematic analysis treats coding as interpretive work and not as measurement.

Early in this process, two broad themes organized our comparisons, suggested in part by the weight of the Features and Lived Experience codes in the corpus. One theme concerned how participants interpreted the functionality presented by the platform, including what a setting appeared to do, how strongly it operated, and how its wording or behavior varied across platforms. The second theme concerned how participants interpreted those same controls through their own routines, relationships, expectations, prior experiences, and concerns. We initially explored the two themes separately, and repeated comparison showed that they were difficult to sustain as distinct patterns. Participants' understandings of screen time, notifications, account privacy, reporting, data access, and deletion moved between what the interface presented and what participants already knew, expected, or needed from social media. We developed \textit{Living With and Around Platform Design} to capture the relationship between platform-provided controls and the experiences participants brought to their interpretation.

A second pattern was identified through cross-case comparison of how participants responded when controls were unfamiliar, confusing, burdensome, or inconsistent with their expectations. We returned particularly to excerpts coded as Expectation, Reflection, Data and Privacy, and Usage, where participants left defaults unchanged, treated additional steps as evidence of safety or care, used familiarity or perceived competence to justify trust, or continued to position the platform as the actor capable of resolving uncertainty. The responses differed in their specific form and did not occur uniformly across participants. Their recurring relationship led us to develop \textit{Teens Defer to the Platform as Authority}, which captures how participants could question particular controls while continuing to treat platform defaults, explanations, designers, and existing configurations as reference points for deciding what a protective setting should mean or how it should work. We refined both themes through continued comparison with the complete dataset, including excerpts that supported, complicated, or contradicted the developing interpretations. We also considered whether an interpretation had emerged organically during task interaction or through interviewer prompting when deciding how broadly a claim could be stated. Through repeated review, memo writing, discussion, and revision, we arrived at the final themes presented in Section~4.

\subsection{Researcher Positionality}
In this study, we acknowledge our interdisciplinary backgrounds and our relationship to teenagers and social media.
Before conducting the study, two researchers were already familiar with privacy and safety tasks on teen social media
accounts through the project framing process and prior exploration of settings across platforms. Our research team
included four members with expertise in education, law, youth studies, security, design, and HCI. These different
perspectives helped us examine how interface design and dark patterns shape privacy and safety settings and influence
how teenagers understand and use social media.

\section{Findings}
In this section, we report on two themes: \textit{Living With and Around Platform Design} describes how participants worked out what a protective setting meant and whether it was useful, drawing on their routines, boundaries, and prior experiences when the setting itself did not tell them. \textit{Teens Defer to the Platform as Authority} describes what happened when a setting was confusing or did not behave as expected, and how participants continued to treat the platform as the reference point for what protection should look like.

\subsection{Living With and Around Platform Design}

In this theme, we examine how participants made sense of platform features intended to support safety, privacy, and control. Protective settings were often softer and less predictable than participants expected, since screen-time reminders could be snoozed, temporary controls allowed participants to step back without leaving, and similar features worked differently across platforms. Everyday routines shaped how participants understood and used features, including how they studied, spent time online, stayed available to friends, and managed interruptions. Questions of privacy and safety were also tied to personal boundaries around unwanted contact, personal information, and visibility. Familiarity could make a feature easier to approach, but clear steps, recognizable risks, and visible outcomes still shaped whether it felt useful or controllable. Across these accounts, participants both accommodated the design in front of them and worked around it, adjusting their expectations to what a setting would do while finding their own routes to the outcomes those settings did not deliver.

\subsubsection{The name of a setting did not predict its strength or its operation}
Among the tasks, screen time most often required participants to infer how a protective feature would intervene, with most expecting monitoring rather than a forced stop. PS\_05, for example, expected [Instagram's daily limit] to arrive as a message: \textit{``maybe they say they'll remind me. So they probably send me a notification.''} PS\_11 kept both outcomes open, describing the feature as something that would \textit{``probably log me out or there might be an alarm or notification, indicating that I've spent so much time on Instagram.''} A firmer expectation emerged among participants who imagined direct enforcement. PS\_02 expected the limit to end the session without asking, saying \textit{``maybe that the application is going to close itself,''} while PS\_09 described TikTok's limit as a lockout: \textit{``maybe it locks the app or when I click on the app it says I have no more time available and tells me I can't watch any videos.''} Against those imagined restrictions, the available protection appeared comparatively soft when PS\_07 snoozed the break reminder during the task and continued to scroll. PS\_06 held the firmest version of the expectation, stating that \textit{``if there's a screen time then it should hold,''} and questioned what the feature accomplished once it did not, asking \textit{``what's the use of taking a break if it doesn't, like, kill the screen or something?''} The distance between a limit that stops the session and one that can be dismissed was therefore something participants could articulate during the task, and not only a mismatch visible to us in comparing their expectations against what the platform delivered.

Beyond screen time, participants approached controls over contact and participation as temporary or reversible rather than final. PS\_05 treated notifications as something to reduce for a time instead of turn off, explaining, \textit{``I usually pause my notifications on Instagram.''} PS\_08 likewise separated deactivation from permanent deletion before attempting the deletion task, saying \textit{``sometimes I deactivate the account without deleting it permanently,''} and locating the value of that distinction in the option to return: \textit{``so that I can go back to it if I want to.''} Platform-to-platform variation, however, made protective features harder to anticipate and their outcomes harder to verify. After reaching Instagram's notification menu with expectations shaped by YouTube, PS\_02 observed, \textit{``there's so much more options than YouTube.''} Uncertainty persisted after action as well; upon completing the YouTube report flow, PS\_11 said, \textit{``I don't know if it's worked \ldots I don't know if that's how it is.''} Ultimately, PS\_09 named the broader difficulty by wishing that platforms \textit{``all had the same privacy settings''} so that knowledge gained on one could guide action on the others. What a feature was called therefore told participants little about what it would do. A screen-time limit might remind, notify, or be postponed. A notification control might pause rather than stop. Deletion might resolve into deactivation. A report might end without any visible outcome. Because the same category behaved differently from one platform to the next, participants could not carry a working understanding from one interface to another and instead worked out each platform independently.

\subsubsection{Features are understood through everyday routines}

Participants interpreted protective settings through routines already in place. PS\_11 captured how the same platform could support schoolwork and pull attention away from it, explaining that \textit{``when I'm giving an assignment and I'm having difficulties in solving them, I go to YouTube and I search for it,''} before adding that \textit{``sometimes I get carried away and I start watching videos.''} PS\_02 looked at how much time they already spent on each platform and set the limit at half that amount: \textit{``I use like almost three hours a day on social media and I spend about an hour, 30 minutes on YouTube as well as Instagram. So maybe I should just use for like half. So like 45 minutes.''} PS\_07 likewise brought an existing effort to limit a familiar habit to the task, explaining, \textit{``I really try to set limits for myself. I spend a lot of time on TikTok.''}

Notification and status settings were also understood through social routines, not only as ways to reduce interruptions. For PS\_10, Instagram's sleep mode did not necessarily stop social media use. Instead, it let friends know that PS\_10 was unavailable and might not reply: \textit{``sometimes you might just leave your data on and they think you're active. But once they see this sleep mode, assume, okay, maybe you're sleeping.''} By contrast, PS\_02 declined to pause notifications because doing so could reduce contact with friends: \textit{``I could get text from my friends and if I were to pause the notifications, I probably wouldn't get that a lot.''} PS\_03 focused on urgency rather than friendship, wanting alerts \textit{``every time''} so that \textit{``I don't miss anything important.''} The same control carried different meaning depending on what a participant was already managing, whether that was attention, availability to friends, or the risk of missing something.

\subsubsection{Privacy and safety are interpreted through personal boundaries around contact, visibility, and (dis)comfort}
Participants often explained privacy through decisions about contact, including who could approach them, whom they knew, and when another person's behavior became uncomfortable. PS\_10 described a range of responses to discomfort, including account privacy, blocking, and muting: \textit{``sometimes I want to put my profile on private. And sometimes I may also want to block some people who are actually making me uncomfortable on social media. And sometimes I just mute people.''} Whereas PS\_10 named several ways to manage unwanted interactions, PS\_05 focused on stopping one recurring form of contact, saying \textit{``I get some, like, unsolicited DMs every time, so I just want to stop it.''} PS\_06 set the boundary earlier by choosing to add only people they knew: \textit{``I wouldn't want things from suggested people or friend suggestions because I like to add people that I know and people that know me.''} Familiarity, rather than access alone, guided whom PS\_06 was willing to add.

Where contact concerned who could reach a participant, visibility concerned how much of them was available to be seen. PS\_11 treated not posting as a way to keep others from knowing about them: \textit{``I'm not much of someone who posts things. So that's to me that's a form of privacy, people not really knowing about me.''} Desired visibility still mattered within these boundaries, even though PS\_08 did not describe follower loss as a privacy problem. Recalling an account change, PS\_08 said, \textit{``I lost a lot of followers. Really disappointed me.''} PS\_08's disappointment shows that visibility could feel valuable even when other forms of contact felt unwanted, so a setting that reduced exposure was not straightforwardly welcome.

Being visible to the platform itself raised a separate concern about what the company collected and observed. PS\_04 narrowed that concern to a single field during the data download task, reacting, \textit{``phone number. No, I don't want them to have my phone number.''} For PS\_04, the broad idea of data privacy became concrete at the point of a phone number. PS\_05 described the platform as an observer rather than a recipient, saying early in the interview that platforms were \textit{``probably looking at some of our records''} and \textit{``must be like spying on me some way.''} Asked whether that concern was part of why they spent less time on social media, PS\_05 answered, \textit{``Yes, exactly.''} The concern shaped how much PS\_05 used these platforms without leading them to leave any of them, since they continued using Instagram, Snapchat, and TikTok throughout the session. Privacy in this account was not a boundary a setting could hold, since the party being kept at a distance was also the party providing the setting.

\subsubsection{Features feel usable when their steps, risks, and outcomes are clear}

Familiar steps helped participants begin a task, but they did not guarantee success or action. PS\_07 linked ease to an action they had completed before, saying \textit{``I think I've done it before, so I think it was easy for me to understand.''} PS\_08 carried knowledge from Instagram into TikTok's deletion flow, reasoning that \textit{``it should be the same as Instagram. I know how to. Let me try that.''} When PS\_08 attempted account deletion on Instagram, however, the expected route did not keep them on track: \textit{``I'll go to account center. I'm sort of looking for the [feature]. I think I made a mistake.''} Knowing where to find a feature also did not make it feel safe to use. Before the session, PS\_05 had looked up account deactivation online but did not complete it: \textit{``I've checked it online before, but these steps online. But I didn't go through with it because I was scared. Scared of losing the account.''} Fear of losing the account pulled PS\_05 back toward the platform, even after they knew the deactivation steps. Note that PS\_05 had questioned the researcher about account deletion because of the unsolicited messages described above, so the platform was already producing a harm they wanted to escape. Losing the account was the larger loss, and PS\_05 stayed. Asked about the 30-day retention period in which the account could still be recovered, PS\_05 described it as \textit{``a good function, because I get to go back''}, reading the platform's continued hold on the account as a service rather than as part of what made leaving difficult.

Participants found less familiar features useful when they could connect them to a clear problem, especially account or data loss. PS\_10 understood the data download feature as a backup against losing access, explaining, \textit{``I wanted to just keep the information in case my account gets disabled.''} PS\_02 imagined the same feature through the possibility of changing their mind after deletion, describing wanting the account back and finding that \textit{``I might not be able to get it back.''} Neither participant had used the feature before, and neither described it in terms of what it produced. Both gave it a purpose by naming something they feared losing. Usefulness in these accounts depended less on whether a feature existed than on whether participants could attach it to a problem they could already picture having.

Whether a setting felt controllable also depended on whether the platform produced the outcome a participant expected. PS\_09 first described control as the ability to choose content and contacts: \textit{``I feel like I have control kind of with what I see, like when I'm watching videos on like TikTok or something and control [of] who I talk to.''} Later in the interview, unwanted content contradicted that earlier account: \textit{``sometimes when I don't want to see something and I see it anyway. I just feel like I don't have control because I didn't choose to see that.''} For PS\_09, control felt present while the feed followed their choices and disappeared when unwanted content appeared. Neither assessment rested on anything PS\_09 could verify, since the only available evidence was what the platform showed next. In the theme that follows, we describe how participants continued to treat the platform as the authority on what a setting meant, even when their own expectations had not been met.

\subsection{Teens Defer to the Platform as Authority}

Participants treated platform design as the main guide to what settings meant and how they should work. Confusing defaults were often left unchanged, while extra steps in reporting and deletion flows could be read as signs of safety, organization, listening, or concern. Trust was invoked in multiple ways, including confidence in the people who built the platform, familiarity with connected services, and a sense that no better alternative existed. Even when a design did not meet their expectations, participants often looked to the platform to fix the problem.

\subsubsection{Defaults and confusing design are often accepted rather than questioned}

Participants often left a platform's default settings in place when they were unfamiliar or difficult to understand. On Instagram notifications, PS\_08 described leaving the settings untouched as an established habit: \textit{``I usually don't touch them. That's what I do on my Instagram. I sort of leave it on default? But yes, this is my settings. I don't change anything here.''} PS\_02 connected the same response more directly to confusion, saying, \textit{``Usually if it's a bit confusing to me, I just sort of leave it the way it's. It was programmed.''} Whereas PS\_08 left the default untouched because doing so was familiar, PS\_02 retained it because confusion made leaving the programmed option in place easier than deciding what to change. Dense or lengthy explanations made that decision even harder. Describing data and privacy messages, PS\_02 said, \textit{``Because sometimes when I see, I get messages, like how we process your data, your privacy cookies and a lot of that professional language is still. It's a bit confusing for me. So sometimes I don't fully understand.''} PS\_05 located the difficulty in the amount of information presented during TikTok's deletion flow, saying, \textit{``I think it's too much information for me to understand.''} PS\_08, by comparison, described avoiding lengthy explanations altogether: \textit{``I usually don't read any long things like that.''} The barrier took different forms, unfamiliar language for PS\_02, too much information for PS\_05, and excessive length for PS\_08, but each reduced how far the participant engaged with the explanation.

Limited understanding could also stop participants from acting even when they were not responding to a specific explanation. PS\_05 described uncertainty as a general reason for leaving features unchanged: \textit{``Honestly, I wouldn't change anything because I really don't know much about it.''} Whereas PS\_05 linked inaction to not knowing enough, PS\_02 described confusion as reaching a point where no next step seemed available: \textit{``Honestly, I'd probably just stay stuck and not do anything.''} A setting that failed to respond as expected could move PS\_02 from leaving it unchanged to withdrawing from the interaction entirely: \textit{``Or I just left the settings as they are and if it doesn't respond, I stop using it.''} PS\_02's response places the cost of an unclear setting on the participant, since the existing setup remains in place while the participant stops trying to make it work.

\subsubsection{Friction is often interpreted as care, organization, or responsible design}

When participants encountered extra steps, they often read them as signs that a platform was acting carefully rather than making a task unnecessarily difficult. PS\_03 connected the steps required to open and maintain an account directly to safety, saying, \textit{``I think all of them are kind of like safe in their own way because go through all these rigorous steps to, to open and, you know, keep your account with them.''} By calling them \textit{``rigorous steps,''} PS\_03 treated the effort required as evidence of safety. PS\_05 gave a more specific explanation after encountering several choices in Instagram's account deletion flow: \textit{``There's so many options. You have to pick different options, like different reasons, before you're able to delete the account.''} The same participant then offered their interpretation of that design, saying \textit{``maybe they just want to make sure that we're pretty\ldots we're very sure before deleting the account.''} Whereas PS\_03 connected a longer process with safety when opening and maintaining an account, PS\_05 understood the added deletion steps as a check that the user was sure before deleting.

Reporting flows gave PS\_05 two other reasons to view extra steps positively. On TikTok, the participant interpreted the report categories as a way to organize complaints: \textit{``I think maybe they just want to classify everything under the right category.''} On Instagram, PS\_05 gave the process a more personal meaning, saying, \textit{``I think it's good. That way you get to explain like the company is really listening to me.''} TikTok's categories made the flow appear organized, whereas Instagram's request for an explanation made the company appear attentive to the person submitting the report. Deletion interfaces prompted similar positive assumptions, although participants focused on different parts of the process. PS\_11 treated Instagram's data download prompt and suggested alternatives as guidance offered before leaving the platform: \textit{``So on this screen [download your data in the account deletion flow on Instagram], they are telling me things to do before I finally deactivate it. So it's saying export my information and then deactivate the account. And they're looking for alternatives for me to keep in contact with the account before I delete it.''} Rather than reading the alternatives as a delay, PS\_11 understood them as ways to preserve access before deletion. PS\_10 focused instead on Instagram's request for a reason: \textit{``Is nice. [Instagram] is trying to ask you why you're trying to delete it. And [Instagram] is trying to just show concern and try to get the reasons for why you're deleting the account.''} PS\_11 saw the export option as a way to preserve access, while PS\_10 read Instagram's request for a reason as an effort to \textit{``show concern,''} giving helpful purposes to steps that also made deletion longer.

\subsubsection{``Necessary Evil'': Trust and deference are shaped by familiarity, perceived competence, and constrained choice}

Participants did not describe trust as a simple belief that platforms were reliable. Familiarity with connected services could make a platform feel trustworthy, but it could also make trust seem unavoidable. Speaking about Instagram, PS\_07 said, \textit{``Instagram? And that's just because I feel like I trust them because I also have a Threads account and a Facebook account. So because it's linked to multiple platforms or accounts, then you feel like it's a trusted source.''} Connections among services reassured PS\_07 because Instagram belonged to a larger group of platforms the participant already used. PS\_04 described a less voluntary relationship with YouTube: \textit{``It has to be\ldots Maybe it has to be YouTube and that's just because of Google. I know that I'm using Google and so I kind of have to trust them because I created my email with them.''} Whereas PS\_07 treated connected accounts as a reason to trust Instagram, reliance on Google for email meant that PS\_04 felt they \textit{``have to trust them.''} Signs of competence offered another basis for trust. PS\_09 read frequent changes to an app as evidence of planning and security: \textit{``I feel like their app is like, it changes a lot. So, it feels like they know what they're doing with it and they have a plan or whatever. And it feels like a place where I can keep my information and not worry about it being stolen or taken.''} PS\_08 placed similar confidence in the designers' choices, saying, \textit{``Because I think maybe it's the best way, the way the creators did it.''} Ease of use added to that confidence: \textit{``Just because I feel like it's easier to use so I trust [Instagram] more.''} PS\_09 associated visible changes with active planning and protection of information, while PS\_08 treated the creators' choices and ease of use as reasons to trust Instagram.

Participants continued to look to the company for solutions even when a platform did not meet their expectations. PS\_09 imagined asking the company to correct a problem but was unsure how: \textit{``I could send them, I don't know if they have, like I could send them an email or whatever saying that they need to fix their app, but I don't really know.''} Although PS\_09 did not know how to contact the company, the response still placed responsibility for fixing the app with its designers. PS\_10 also waited for platform action: \textit{``I feel okay, maybe they might adjust their settings and I might actually want to use the app again. So that's why I haven't deleted my account.''} Whereas PS\_09 imagined requesting a fix, PS\_10 kept the account because a future change might make the app acceptable again. Limited alternatives could also keep participants attached without full trust. PS\_02 said, \textit{``I don't fully trust them. But, you know, Instagram is like a necessary evil. I know that. I just feel like maybe they're the best of the bunch, so I have no option really.''} Calling Instagram a \textit{``necessary evil''} separated continued use from confidence because PS\_02 remained with the platform only when the alternatives seemed worse. PS\_05 offered a contrast rooted in a specific negative experience: \textit{``Because I\ldots I don't know. I just don't have a good vibe with [Twitter]. I had an account, and they blocked it, and the reason they gave me was not really good. So I just feel like I don't really trust them.''} PS\_02 remained with Instagram because no better option seemed available, whereas PS\_05's blocked account and the platform's poor explanation gave them a specific reason to withhold trust. In the end, the platform remained the central actor against which their expectations were formed and evaluated.

\section{Discussion}
Across our two themes, we have shown that participants construct the meaning of privacy and safety controls during use rather than receiving that meaning from the interface. Platforms presented protection that was often soft, reversible, and worded differently from one service to the next, and participants worked out what each control meant through the routines, boundaries, and prior experiences they brought to the task. Repeated encounters between platform cues and lived experience could then lead participants to leave confusing defaults in place, to read the effort a flow demanded as evidence of care, and to treat the platform as the authority on what a protective setting should do. We build on these findings in two ways in this section. Section 5.1 examines how participants could recognize where a control failed, and whose interests that failure served, without that recognition altering how they used or interpreted the platform. Section 5.2 turns from the individual encounter to the conditions surrounding it, where we argue that evaluating a control feature by feature cannot capture an ecology in which engagement-oriented strategies recur across platforms while the work of managing protection does not transfer with them. We further argue there that privacy and safety controls depend on different platform functions once a teenager has acted, and that responsibility for whether protection follows should be allocated asymmetrically between teenagers and platforms.

\subsection{``They Want You to Stay'': Between Recognition and Deference}

When PS\_06 asked \textit{``What's the use of taking a break if it doesn't, like, kill the screen or something?''}, the question exposed both the weakness of the screen-time reminder and the participant's awareness of how little control it provided. The same participant identified the platform's incentive directly, observing that the limit was \textit{``very easy to ignore''} because \textit{``they want you to stay in the platform.''} PS\_09 described unwanted content as the moment choice collapsed, explaining that \textit{``I don't want to see something and I see it anyway,''} while PS\_02 said plainly that they did not fully trust Instagram. Participants could therefore recognize where protective controls failed and whose interests those failures served. Yet their behavior rarely converted that clarity into refusal, adjustment, or exit. During the task itself, one participant snoozed the screen-time limit (PS\_07), another expected to ignore the reminder and continue scrolling (PS\_05), and the participant who distrusted Instagram remained on the platform (PS\_02). Their recognition remained diagnostic rather than disruptive, exposing how platforms shaped their choices without providing a durable basis for resisting those choices or changing their relationship with the platform.

The gap between recognition and resistance was sustained by how participants interpreted the friction surrounding privacy and safety controls. Across notification settings, reporting, and account deletion, participants repeatedly treated the effort demanded from them as evidence of effort being made for them. Multi-step authentication made platforms appear \textit{``safe in their own way''} because users had to complete \textit{``all these rigorous steps''} (PS\_03), allowing procedural difficulty to communicate protection rather than obstruction. Four other participants gave protective motives to steps that lengthened a task, describing deletion prompts as checks that a user was certain, reporting categories as a way of organizing complaints, requests for a reason as concern, and export options as help preserving access (PS\_05, PS\_10, PS\_11). In the vocabulary of dark patterns scholarship, these features represent obstruction, forced action, and sneaking \cite{Gray2024-rf, Gray2018-it}, while participants described them through the language of thoroughness, organization, and care. Those competing vocabularies located responsibility in different places. Obstruction draws attention toward the platform's attempt to constrain a choice, whereas thoroughness allows the same constraint to appear as evidence that the platform is exercising responsibility on the user's behalf. Participants' language moved attention away from retention incentives and toward protective motives, making additional steps signify safety, repeated questions signify certainty, detailed categories signify attentiveness, and alternatives to deletion signify concern for what users might lose. For teenagers confronting billion-dollar platforms that design the choice, the friction surrounding it, and the explanation available for interpreting it, visible complexity can become a powerful signal of institutional competence. Under these conditions, design does not need to conceal how it shapes users' decisions because its visible burden becomes what teenagers cite as the reason to trust it.

Interpreting interface friction as care did not mean that participants trusted platforms without reservation. Participants recognized that platforms could frustrate, manipulate, or fail them, yet they continued to rely on platform defaults and explanations when deciding what a protective setting was supposed to do. Prior work has established that teenagers can identify privacy dark patterns when directed to examine them \cite{Kelly2025-re, Sanchez-Chamorro2024-hv, Schafer2024-gu}, while our findings show that recognition can emerge without prompting and still leave the relationship with the platform unchanged. Nagy and Neff describe how expectations shape what users understand a technology as permitting \citep{Nagy2015-fl}, and research on folk theories explains how people construct accounts of intention when system behavior remains opaque \cite{DeVito2021-pv, Eslami2016-hn, Mayworm2024-sl}. Participants relied on that interpretive logic when uncertainty made the platform's judgment appear more reliable than their own. One participant accepted default notification settings because \textit{``maybe it's the best way ... the creators did it''} (PS\_08), and another left confusing settings untouched for the same reason (PS\_02). Zhang and colleagues show that trust and distrust can coexist when users evaluate different aspects of the same system \cite{Zhang2022-ky, Zhang2024-ln}, and that overlap appeared when PS\_02 distrusted Instagram while still treating its programmed settings as the safest available judgment. Platform authority persisted not because participants trusted the platform's motives, but because they had no independent standard for deciding whether a setting protected them, manipulated them, or simply failed to work. Meanwhile, the scarcity of alternatives extended beyond interpretation into recourse, leaving participants dependent on the same companies whose decisions they questioned. When PS\_09 imagined responding to a platform that had failed them, the only available action was to \textit{``send them an email or whatever saying that they need to fix their app,''} followed immediately by \textit{``but I don't really know.''} Another participant retained an unused account because the platform \textit{``might adjust their settings''} and become desirable again (PS\_10), while PS\_04 described trusting YouTube less as a choice than as an obligation created by already belonging to Google's ecosystem. Even prior harm did not necessarily loosen that dependence, as repeated spam prompted PS\_05 to investigate deactivation before fear of \textit{``losing the account''} stopped the process. Such dependence gives practical force to the description of platforms as a necessary evil because the platform remains the source of the problem, the dominant authority for interpreting the problem, and the only imaginable venue for repairing it. Research on digital resignation describes concern that persists alongside a perceived inability to change institutional practices \citep{Draper2019-ip, Hargittai2016-xn}, but participants' responses reflected deference more than resignation because they continued to expect protection while believing that only the platform they already doubted could provide it. Design and policy responses that focus on making settings easier to see and understand may therefore target the wrong problem. More elaborate and visible protective flows can appear to teenagers as evidence of diligence rather than obstruction, allowing interventions that expose manipulative patterns to reinforce the authority they are intended to challenge. Evaluations should instead examine whether teenagers can verify that a protective action took effect, whether its protection persists without continued management, and whether they can challenge a platform's decision through an independent authority. Protective design should be judged by the power it gives teenagers rather than by how clearly a billion-dollar platform explains the choices it created, ensuring that young users can confirm that protection works, contest platform decisions, and obtain support beyond the platform's own judgment about what is best for them.

\subsection{The Limits of Feature-Specific Evaluation of Privacy and Safety Controls in a Platform Ecology}

Providing a control did not determine the protection teenagers could actually obtain. As Section 5.1 showed, participants could identify weak or confusing controls and recognize that platform interests did not always align with their own, yet they still relied on platform defaults and explanations to determine what protection should look like. Where protection depended on a setting, participants were required to make consequential decisions through categories, options, and explanations of system behavior supplied by the same platforms whose practices they were trying to manage. Protective defaults can reduce that burden when a platform configures protection before a teenager encounters a setting rather than waiting for the teenager to seek it out, but offering a control does not by itself establish meaningful agency or complete the platform's protective work. The available choice remained partial because platforms determined which options existed, how those options were framed, which defaults applied, and what occurred after a selection, while participants often could not observe or verify the resulting data practice or safety response. That partiality was ecological as well as local, because engagement-oriented strategies recurred across the platforms participants used while the interpretive work of managing protection did not transfer with them. Participants wanted consistent settings across platforms so that knowledge gained on one could guide action on another, carried expectations from one deletion flow into another and lost the path, and described connected accounts and shared infrastructure as reasons that trust in one service extended to another. Comparing a feature across platforms, whether by researchers or by teenagers themselves, treats the control as separable from these conditions and from the platform authority that structures its use. Habib and Cranor's emphasis on awareness, comprehension, effort, reversibility, and exposure to nudging identifies important qualities of a control \cite{habib2022evaluating}, while participants' accounts show how those qualities depended on interactions among defaults, explanations, system behavior, everyday routines, and the wider environments in which controls operate.

Privacy and safety meet in the gap between a teenager's action and the protection that follows. The platform function on which that protection depends, however, differs across the two domains. Privacy controls require teenagers to make judgments about collection, inference, sharing, and retention that occur beyond ordinary observation, leaving them dependent on platform explanations and faithful implementation to know whether a selection changes the underlying data practice. Safety controls create a procedural dependence because blocking or reporting initiates a request for protection while the platform classifies the incident, reviews evidence, chooses a remedy, and determines whether enforcement follows. A combined privacy and safety framework can support future research without treating the domains as interchangeable, directing attention to the point at which a visible user action passes into an organizational process that the user cannot independently inspect or complete. Privacy analysis can examine whether a selected setting changes collection, inference, sharing, or retention as represented, while safety analysis can examine whether a report, block, or request produces review, enforcement, and protection through procedures that users can understand and challenge.

In this analysis, responsibility first denotes an organizational obligation to avoid foreseeable privacy and safety harms and to ensure that protective actions work as represented, while its governance and regulatory meaning concerns how those obligations are assigned, monitored, and enforced. Participants' sensemaking shows that evaluating controls only in terms of whether teenagers can locate and complete them captures the visible interaction but not the interpretive work through which they assessed what protection a control could provide. Participants also had to infer which options the platform had already judged to be protective, what safeguards operated behind the interface, and which actor could change the conditions producing risk. Opposing interpretations of the same control assigned responsibility in different directions, with platforms treating the availability of a control as fulfillment of their obligation while participants understood the same control as evidence that the platform had already evaluated the option and would make it effective. Teenagers contribute contextual knowledge, expressed preferences, and reports of harm, while platforms control the defaults, explanations, implementation, and moderation procedures that determine what those contributions can accomplish. Responsibility should therefore be allocated asymmetrically, with platforms carrying obligations to test implementations, monitor outcomes, correct failures, and respond when reports or protective selections reveal a problem. Richards and Hartzog's account of loyalty provides a normative basis for this allocation because a platform that designs the available options and controls their downstream effects holds knowledge, authority, and capacity that users do not possess \cite{richards2021duty}. Connecticut's statutes provide a related institutional example by requiring secure and reliable account-removal procedures, prescribed response periods, cessation of personal-data processing after deletion subject to statutory exceptions, reasonable-care duties, and assessments of reasonably foreseeable risks to minors \cite{ctgs2026minors}.

\section{Implications and Future Work}

Platforms can present a control as protective while leaving it hard to reach or easy to reinterpret in ways that serve platform goals, which means that assessment cannot stop at whether a setting exists. Building on work connecting HCI scholarship to platform governance and policy~\cite{Gairola2025-iy,Yang2024-mu}, we propose that evaluation instead ask whether a teenager can form a clear protective intention, locate the relevant setting without expert guidance, carry a protective change through to completion without undue friction or reversal, and understand the outcome well enough to trust that it will last. Our study offers a methodology for treating these as empirical questions rather than theoretical ones, and our data suggest that current platforms fall short on each dimension for at least some settings. Studies should independent completion, teen-reported confidence, and the durability of a protective change as evidence relevant to whether a platform is genuinely providing protective tools rather than merely hosting them~\cite{Yang2024-mu,Jackson2014-wc}.

Future work should extend this account in three directions. (1) Our study captures one task-based session, and whether the difficulty of reaching a setting translates into avoidance over time is unresolved. (2) Participants held varied and often inaccurate mental models of what protective features do, expecting screen-time limits to lock an application or reporting to remove content immediately, and these mismatches follow predictably from soft, reversible, and ambiguously labeled design. Whether interventions such as honest framing of what a limit will and will not do, or visible confirmation that a protective action has taken effect, can close that gap without being absorbed into the interpretive pattern documented in Section~5.1 is worth testing directly. (3) What our method can show is bounded by what a task-based session makes visible. Participants worked through tasks in an interview setting and not in the course of their own use, and some tasks were performed on accounts that did not carry their followers, message histories, or years of accumulated activity. Participants' explanations of what a setting meant are therefore situated in the session, and a teenager reasoning about deleting an account with no social stakes attached is not reasoning under the conditions that make deletion consequential. We also explained or located features at points, so some interpretations reflect understanding built during the session rather than knowledge brought to it. We coded those interpretations as prompted, but the boundary between spontaneous and supported understanding is not always sharp.

\section{Conclusion}
In this paper, we report on our analysis of think-aloud sessions with 11 teenagers completing privacy and safety tasks on social media platforms they already used. Our findings show that participants worked out what a control meant through their own routines and prior experiences, accommodating protections that were softer than they expected, and that recognizing a control as weak did not keep them from treating the platform as the authority on what protection should look like. We argue that evaluating controls one feature at a time cannot establish whether teenagers are protected, and that responsibility belongs with the platforms that design the options and control what follows from them. Assessment should instead ask whether a protective action can be
confirmed, whether it holds without continued management, and whether teenagers can contest a platform's decision elsewhere.

\bibliographystyle{ACM-Reference-Format}
\bibliography{sample-base-cleaned-with-urls}

\appendix

\begin{table*}[t]
\centering
  \Description{Codebook with four code groups, each row giving a code name, its definition, an example excerpt, and an application count. The structural group holds Platform and Task, each applied 326 times. The sensemaking group holds Features at 177, Lived Experience at 263, Usage at 83, Data and Privacy at 59, and Comparison at 22. The reflection and expectation group holds Reflection at 111 and Platform Expectation at 21. The interpretation context group holds Organic at 162 and Prompted at 36. Lived Experience and Features carry the most applications, and Platform Expectation and Comparison the fewest.}

\caption{Code definitions used in the sensemaking analysis.}
\label{tab:codebook}
\small
\renewcommand{\arraystretch}{1.25}
\setlength{\tabcolsep}{0pt}
\begin{tabular}{@{}>{\raggedright\arraybackslash}p{0.18\linewidth}>{\raggedright\arraybackslash}p{0.72\linewidth}>{\raggedleft\arraybackslash}p{0.06\linewidth}@{}}
\toprule
\parbox[c][15pt][c]{\linewidth}{\ \textbf{Code}} &
\parbox[c][15pt][c]{\linewidth}{\textbf{Definition and example excerpt}} &
\parbox[b]{1.6cm}{\centering\textbf{\textit{Number of codes}}}\\\midrule

\grouphead{Structural codes}{Applied to every coded moment to locate it within a specific platform and task.} \\
Platform & Which platform the moment occurred on (Instagram, TikTok, Snapchat, or YouTube). & 326 \\
\addlinespace[3pt]
Task & Which privacy or safety task the moment occurred in (notifications, screen time, private account, reporting, data download, or account deletion). & 326 \\
\addlinespace[5pt]

\grouphead{Sensemaking}{Moments where participants made sense of a feature while performing a task, including what the feature meant, how it worked, and how it related to their own use of the platform.} \\
Features & Participant works out what a feature does or how it works during the task itself.\newline \textit{``I'll probably go to my profile. A screen time, that's like a time limit, right? I think I should know how to do that. So there's the hamburger menu settings.''} (PS\_01) & 177 \\
\addlinespace[3pt]
Usage & Warm up or contextual talk about platform habits and familiarity.\newline \textit{``To check out new music, I like that. Also to like keep up with celebrity news''} (PS\_02) & 83 \\
\addlinespace[3pt]
Data and Privacy & Broader talk about privacy, safety, data use, trust, or control.\newline \textit{``[Privacy and safety online is] preventing my account from getting hacked.''} (PS\_11) & 59 \\
\addlinespace[3pt]
Lived Experience & Participant connects a feature to their own routines, harms, goals, or prior experiences.\newline \textit{``Most times when I'm giving an assignment and I'm having difficulties in solving them, I go to YouTube and I search for it and I try to watch videos on how to overcome them.''} (PS\_11) & 263 \\
\addlinespace[3pt]
Comparison & Participant compares features across platforms, or transfers understanding from one platform to another.\newline \textit{``The account deletion process is what I did on TikTok that I wanted to do on Instagram.''} (PS\_07) & 22 \\
\addlinespace[5pt]

\grouphead{Reflection and expectation}{Moments where participants reflected on the feature after attempting the task, including broader judgments about the platform, the experience, and what they thought should have happened.} \\
Reflection & Follow up interview talk where participants reflect on tasks, evaluate the platform, or explain what the experience changed for them.\newline \textit{``Well, maybe it allows you to choose what you want to see and also it just allows each you decide which notifications you want to get.''} (PS\_02) & 111 \\
\addlinespace[3pt]
Platform Expectation & Participant states what the setting, feature, or platform ought to do.\newline \textit{``If there's a screen time then it should hold.''} (PS\_06) & 21 \\
\addlinespace[5pt]

\grouphead{Interpretation context}{Contextual markers recording how an interpretation entered the interview. These codes do not represent task completion outcomes (was not included in the thematic analysis).} \\
Organic & The interpretation emerged without an interviewer probe directing the participant toward that interpretation.\newline \textit{``So it's on the private account. Now let me switch to public. Then I switch it back to private.''} (PS\_07)  & 162 \\
\addlinespace[3pt]
Prompted & The interpretation emerged following an interviewer probe, clarification, navigation assistance, or request for further explanation. \textit{``I'll go to account center. I'm sort of looking for the future. I think I made a mistake.''} (PS\_08) & 36 \\
\bottomrule

\end{tabular}
\caption{Code definitions used in the sensemaking analysis. Analytic codes were nonexclusive, and counts represent code applications rather than unique excerpts.}
\end{table*}

\end{document}